\documentclass{article}

\usepackage{arxiv}

\usepackage[utf8]{inputenc} 
\usepackage[T1]{fontenc}    
\usepackage{hyperref}       
\usepackage{url}            
\usepackage{booktabs}       
\usepackage{amsfonts}       
\usepackage{nicefrac}       
\usepackage{microtype}      
\usepackage{lipsum}
\usepackage{graphicx}
\graphicspath{ {./images/} }
\usepackage{physics}
\usepackage{braket}
\usepackage{algorithm}
\usepackage{algorithmic}
\usepackage{multirow}
\usepackage{mathtools}
\usepackage{diagbox}
\usepackage{xcolor}
\usepackage{times}
\usepackage{epsfig}
\usepackage{graphicx}
\usepackage{amsmath}
\usepackage{amssymb}
\title{QFCNN: Quantum Fourier Convolutional Neural Network}

\author{
 Feihong Shen \\
  Jilin Univeristy\\
  \texttt{shenfh5517@mails.jlu.edu.cn} \\
   \And
 Jun Liu\\
  Singapore University of Technology and Design\\
  \texttt{junliu@sutd.edu.sg} \\

}

\begin{document}
\maketitle


\begin{abstract}
The neural network and quantum computing are both significant and appealing fields, with their interactive disciplines promising for large-scale computing tasks that are untackled by conventional computers. However, both developments are restricted by the scope of the hardware development. Nevertheless, many neural network algorithms had been proposed before GPUs become powerful enough for running very deep models. Similarly, quantum algorithms can also be proposed as knowledge reserve before real quantum computers are easily accessible.
Specifically, taking advantage of both the neural networks and quantum computation and designing quantum deep neural networks (QDNNs) for acceleration on Noisy Intermediate-Scale Quantum (NISQ) processors is also an important research problem.
As one of the most widely used neural network architectures, covolutional neural network (CNN) remains to be accelerated by quantum mechanisms, with only a few attempts have been demonstrated. In this paper, we propose a new hybrid quantum-classical circuit, namely Quantum Fourier Convolutional Network (QFCN). 
Our model achieves exponential speed-up compared with classical CNN theoretically and improves over the existing best result of quantum CNN.
We demonstrate the potential of this architecture by applying it on different deep learning tasks, including traffic prediction and image classification.
\end{abstract}

\section{Introduction}
Existing works \cite{biamonte2017quantum,dunjko2018machine} have attempted to incorporate the classical machine learning algorithms and quantum computation techniques, which yields quantum machine learning frameworks.
It is also promising for deep convolutional neural network (CNN) to be leveraged with quantum methods towards achieving quantum convolutional models, thanks to the following reasons.  
First, CNNs manipulate data in the manner of performing matrix operations in high dimensional vector space, while quantum computation also conducts matrix operations in high dimensional vector space~\cite{biamonte2017quantum}. Second, most of the deep CNN models require large datasets for parameter optimization, and often need heavy computation for both the training and testing. Meanwhile, based on special quantum information properties named entanglement and superposition, quantum computers can make exponential number of calculations in parallel. This mechanism, known as quantum speedup, can accelerate some learning algorithms to surpass their classical counterparts based on mathematical proofs \cite{kerenidis2019q,lloyd2013quantum,kerenidis2020quantum,wiebe2014quantum}.

Two decades ago, scientists began to implement quantum algorithms on real quantum machines. With the continuous development of quantum hardware, scientists suggest that human beings will live in Noisy Intermediate-Scale Quantum (NISQ) era in the future, and everyone is expected to have a quantum computer with 50-100 qubits, which could be used to perform some special tasks that surpass the capabilities of today's classical computers \cite{tannu2018case,preskill2018quantum}.

Recently, deep neural networks become quite popular and have been widely applied in computer vision, natural language processing, and many other areas, thanks to the advancement of deep learning techniques \cite{redmon2016you,karras2019style,devlin2018bert}. As one of the most important deep learning techniques, the convolution module has been used in many network architectures, and many state-of-art models for handling different types of tasks contain convolutional layers \cite{bochkovskiy2020yolov4,he2016deep,huang2017densely,murez2018image}.

Given the aforementioned background of quantum computation and flourish CNN fields, it becomes natural that we get interested in integrating the potentials and advantages of these two kinds of techniques, i.e., quantum computing and CNNs, to handle specific convolution-based learning tasks that need continuously increasing computation capabilities. However, the direct convolution of quantum state has been shown to be inaccessible by previous studies ~\cite{lomont2003quantum}. In this paper, we propose a hybrid circuit with quantum Fourier transform to achieve deep convolutional models with quantum.

The major blocks in CNNs are the convolutional operations, i.e., the product of the kernel and the input data that can have various dimensions. This process often accounts for a large proportion of the computation cost and time consumption in the whole network. Another approach that can be used to efficiently implement the convolutional operation is transforming the kernel and data to Fourier basis and then performing inverse transform based on their product. Some of the classical networks like graph convolutional networks (GCNs) have used the approximation of this method to implement convolution~\cite{kipf2016semi,pratt2017fcnn,rippel2015spectral}. 
Inspired by this, to achieve quantum convolution, here we propose a novel quantum Fourier convolutional network (QFCN) to speedup the CNN with quantum Fourier transform (QFT), which could be exponentially faster than the classical Fourier transform for convolutions. We circumvent the inaccessibility issue \cite{lomont2003quantum} by adopting part quantum circuit of windowed Fourier transform \cite{ma2019windowed}. We also propose a hybrid quantum-classical circuit to avoid some of the quantum noise and also to take advantage of classical computers that are able to perform rapid iterations for model optimization.
We demonstrate the potential of our network on the traffic prediction task with spatio-temporal graph convolutional models and the image classification task with 2D convolutional networks.

{\bf Motivations and contributions.} 
We all know the importance of convolution to the deep learning field. If we want to speed up the training of nowadays’ networks which mostly build upon a bunch of convolutional layers, the quantum convolutional layer and its optimization will be inevitable problems. There are only two architectures of convolutional layer being proposed\cite{kerenidis2019quantum,cong2019quantum} . However, due to the horrific time assumption of quantum stimulation, the results of these two papers are based on toy experiments and their network only contain two quantum convolutional layers. Unfortunately, with the increase of the layer number, the gradient of these two structures may lead to Barren plateaus~\cite{mcclean2018barren} in training. So, it is essential to proposed another quantum convolution method rather than simply transferring conventional multiply accumulation (MAC) to quantum mechanism like previous works.

The contributions of this paper are summarized below. To the best of our knowledge, we are the first to propose a hybrid quantum-classical circuit to speed up CNN with a parametric quantum circuit (PQC), and exploit quantum Fourier transform for the trainable convolution models. Specifically, our PQC can be inserted into different deep networks to replace the convolutional layers.
Besides,
we also introduce the quantum optimization algorithm (back-propagation) for our designed network. 

\section{Related Work}
{\bf Quantum machine learning.}
Quantum machine learning algorithms can be roughly classified into two groups \cite{biamonte2017quantum,preskill2018quantum}. The first group is the algorithms proposed before the concept ``NISQ'' have been published, such as quantum PCA \cite{lloyd2014quantum} and quantum SVM \cite{rebentrost2014quantum}, and most of them are still in the developing period. The second group is mainly the deep models with quantum network layers. Previous research works \cite{zhao2019qdnn,farhi2018classification,chen2018universal} have shown the feasibility of training the networks on NISQ devices. There are several quantum deep learning architectures \cite{kerenidis2019quantum,lloyd2018quantum}, such as 
quantum recommendation systems \cite{kerenidis2016quantum}, quantum generative adversarial networks \cite{dallaire2018quantum,lloyd2018quantum}. 
Some models like graph networks \cite{melnikov2019predicting} and recurrent neural networks \cite{hibat2020recurrent} have also been inspired by quantum computing.

{\bf Convolution.} As a popular neural network model, CNN usually contains four types of layer: the convolution layer, the activation function, the pooling layer and the fully connected (FC) layer. 
Recently, there have been a few attempts \cite{kerenidis2019quantum,cong2019quantum} for designing quantum convolution models, and they mainly focus on   
using the quantum gate to simulate the
layers in traditional CNNs. 
Differently, in our QFCN model, we exploit discrete quantum Fourier transform for the learnable convolutional modules, which still
enables us to control the qubits with the number of kernel size. Besides, our designed quantum circuit makes it convenient to prepare the input signal to be involved in convolution. Also note that besides deep CNN, there are a few attempts focusing on pure quantum convolution without network architecture and learnable parameters \cite{ma2019windowed}, and our method is also different from them, since we aim to design a quantum Fourier convolutional neural network with trainable parameters, and the quantum optimization algorithm via back-propagation is also introduced in this paper to learn the parameters for our QFCN. 

{\bf Hybrid quantum-classical algorithm.} While quantum computers still face considerable technological challenges, many quantum algorithms have daunting resource requirements~\cite{mcclean2016theory}. To handle this problem, some quantum algorithms \cite{peruzzo2014variational} connected with classical computers are designed, which are often called hybrid quantum-classical algorithms. Besides saving quantum resources like quantum gates, hybrid quantum-classical algorithms also show other advantages. For instance, classical computers are able to perform rapid iteration. Thus some hybrid schemes make use of this advantage to optimize the parametric quantum circuit~\cite{peruzzo2014variational}. Here the hybrid framework is exploited to design our QFCN model, considering its advantages.

\section{Preliminaries and Notations}
\label{section:preliminary}

In this section, we discuss the basics of quantum computing and quantum information as the context of this paper, to facilitate the subsequent sections that introduce our proposed quantum Fourier Convolutional Network with the hybrid quantum-classical circuit. 

\begin{figure}[t]
\begin{center}
{\includegraphics[width=\linewidth,trim={0cm 2cm 0cm 0cm},clip]{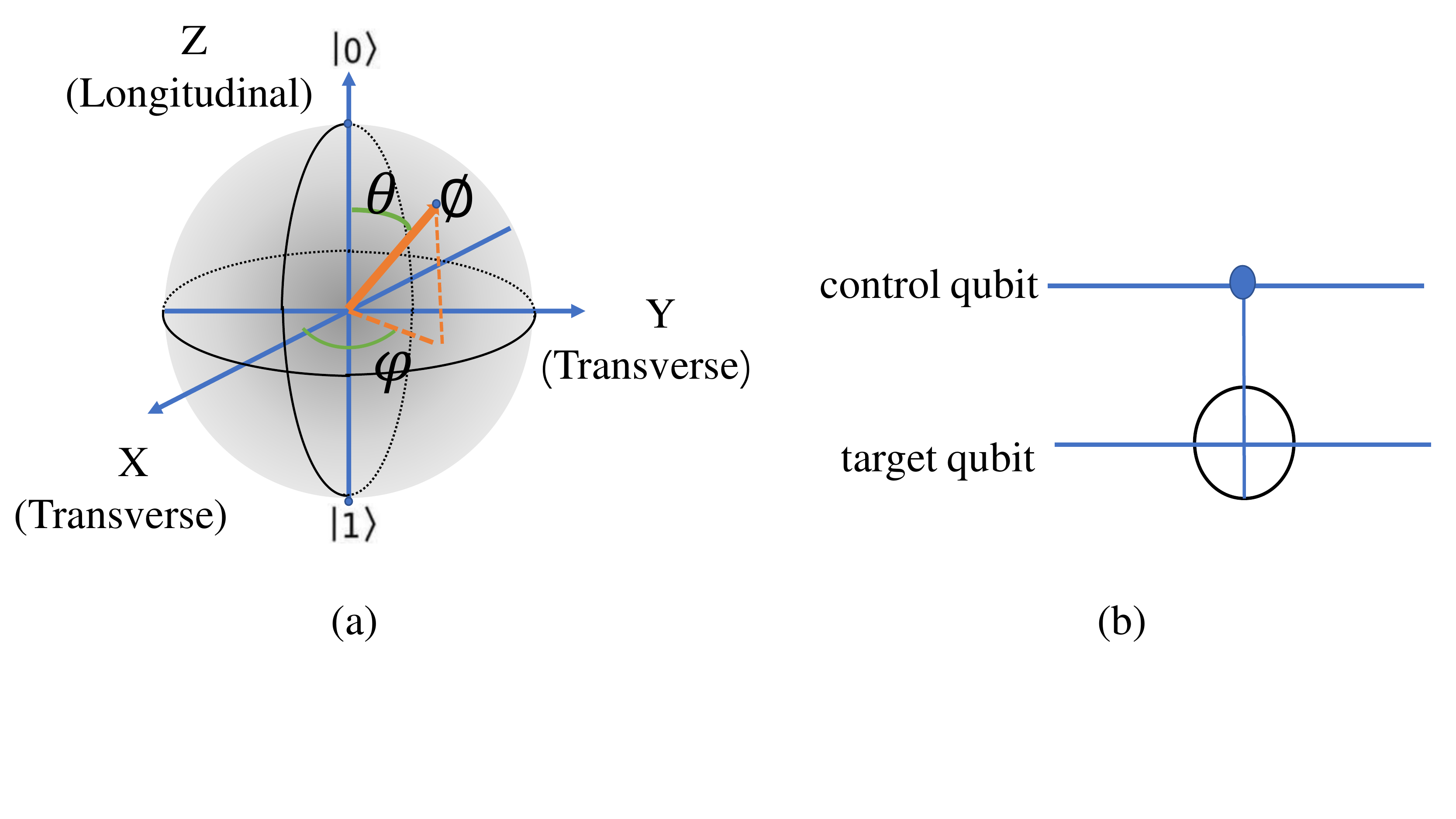}}
   
\end{center}
   \caption{(a) Bloch sphere suggests another representation of a single qubit: $|\phi\rangle=\cos \frac{\theta}{2}|0\rangle+e^{i \varphi} \sin \frac{\theta}{2}|1\rangle$, where $\theta \in[0, \pi] \quad$, and $\quad \varphi \in[0,2 \pi]$. (b) Quantum circuit representation for the controlled not gate.}
\label{fig:bloch}
\label{fig:onecol}
\end{figure}

{\bf qubit.}  Analogous to the bit in classical computers, qubit is a fundamental concept in quantum computing. Actually, qubits can represent both physical objects and abstract mathematical objects. As a physical object, the most important properties for qubits include the superposition for single qubit and entanglement of multi-qubits. 
\textit{(1) Superposition.}
Generally, a classical bit can only be in either state 1 or 0, while in quantum linear algebra, we analogize these two basic states to $|1\rangle$ and $|0\rangle$, under the Dirac notation. A qubit is expressed by a linear combination of the two quantum states, denoted by:
\begin{equation}
 |\phi\rangle=\alpha|0\rangle+\beta|1\rangle \label{1}
\end{equation}
where $\alpha$ and $\beta$ are probability amplitudes. The superposition can be viewed intuitively from the Bloch Sphere, as shown in Figure \ref{fig:bloch}a. The positive direction of the z-axis represents $|0\rangle$, and the negative direction represents $|1\rangle$. $\theta$ and $\varphi$ are the phases off the z-axis and x-axis, respectively. Traversing all $\theta$ and $\varphi$ values, the qubit arrow forms the Bloch sphere. After the observation of this single qubit, its state will collapse into $|1\rangle$ with possibility $|\beta|^{2}$, and $|0\rangle$ with possibility $|\alpha|^{2}$, where $|\alpha|^{2}+|\beta|^{2}=1$. \textit{(2) Entanglement.} The quantum entanglement is a relation between qubits. Assume we have two qubits with unknown states to be entangled, if we observe the first qubit and get its state, the second qubit, at the same time, will collapse into the same state regardless of the distance. 
Any entangled state cannot be represented by the production of two single qubit state, for instance the state $|\phi\rangle=\frac{1}{\sqrt{2}}|a a\rangle+\frac{1}{\sqrt{2}}|b b\rangle$, cannot be decomposed by:
\begin{equation}
|\phi\rangle=\left(c_{1}|a\rangle+c_{2}|b\rangle\right) \otimes\left(c_{1}^{\prime}|a\rangle+c_{2}^{\prime}|b\rangle\right)=\frac{1}{\sqrt{2}}|a a\rangle+\frac{1}{\sqrt{2}}|b b\rangle \label{2}
\end{equation}
where $\otimes$ means tensor product operation. It is obvious that we cannot find the solution of $c_{1} c_{1}^{\prime}=c_{2} c_{2}^{\prime}=\frac{1}{\sqrt{2}}$ and $c_{1} c_{2}^{\prime}=c_{2} c_{1}^{\prime}=0$. 
As a mathematical object, we use linear algebra in quantum way to represent the state of the qubit. Usually, we identify the state $|1\rangle$ with the vector $[0,1]^{T}$ and state $|0\rangle$ with the vector $[1,0]^{T}$.
Moreover, according to quantum linear algebra, we can represent the state $|\psi\rangle$ in binary, such as : $j=j_{1} 2^{n-1}+j_{2} 2^{n-2}+\ldots+j_{n} 2^{0}$. We can also represent the binary fraction $j_{l} / 2+j_{l+1} / 4+\ldots+j_{m} / 2^{m-l+1}$ by $0 . j_{l} j_{l+1} \ldots j_{m}$.

{\bf Quantum gate.} Just like logic gates in classical computer, quantum gates are operators that can manipulate the state of qubits. In physics, the quantum gate can be regarded as the remove of qubit arrows on Bloch sphere from one point to another. In mathematics, quantum gates can be regarded as unitary matrices operating on the vectors of quantum states. There are two kinds of quantum gates: the single-qubit gate and the multi-qubit gate. All single-qubit gate can be written as the 2$\times$2 unitary matrix. For instance, the Hadamard gate is a frequently used single-qubit quantum gate, that allows to produce superposition of states from the basic states. We can represent the Hadamard gate with the unitary matrix:
$\frac{1}{\sqrt{2}}\left[\begin{array}{cc}1 & 1 \\ 1 & -1\end{array}\right]$. We can apply this gate on $|0\rangle$ state:
\begin{equation}
H|0\rangle=\frac{1}{\sqrt{2}}\left[\begin{array}{cc}1 & 1 \\ 1 & -1\end{array}\right]\left[\begin{array}{l}1 \\ 0\end{array}\right]=\frac{1}{\sqrt{2}}\left[\begin{array}{l}1 \\ 1\end{array}\right]=\frac{|0\rangle+|1\rangle}{\sqrt{2}}
\end{equation}
The above equation explains why quantum gates are unitary matrices, i.e., the sum of amplitudes' square of the output state should be 1. 

Another gate in quantum circuits is the multi-qubit gate, also called controlled gate.  The input qubits can be divided into the target qubit and the control qubit. Suppose $U$ is a multi-qubit gate. If the control qubit is set, then $U$ is applied to the target qubit. Otherwise, the target is left alone. This means: $|a\rangle|b\rangle \rightarrow|a\rangle U^{a}|b\rangle$. The controlled$-NOT$ gate is one of the widely-used multi-qubit gates. If the control qubit is $|1\rangle$, then the qubit arrow on Bloch sphere will be flipped 180 degree. The controlled$-NOT$ gate is illustrated in Figure \ref{fig:bloch} (b).

{\bf qRAM.}
As a fundamental device in the computer, the random access memory (RAM)~\cite{reed2001molecular} stores information in its memory cell. The RAM ascribes a unique address to the memory cell. If we want to read out information, we need give the RAM an address and move the information into the register. The quantum computer also has the counterpart equipment, namely, the quantum random access memory (qRAM). A qRAM can be seen as a RAM working in a quantum way. There are many architectures of qRAM, and most of them are based on a promising architecture, called bucket-brigade for qRAM~\cite{park2019circuit,giovannetti2008quantum,arunachalam2015robustness}. For simplicity, in this paper we use ``qRAM'' to denote the ``qRAM with the bucket-brigade architecture''. We input the address of the information in quantum state, and get the data in quantum state as the output~\cite{moreno2020circuit}:
\begin{equation}
\sum_{i} \alpha_{i}|i\rangle_{d i r}|0\rangle_{d a t} \quad \stackrel{\text { qRAM }}{\longrightarrow} \sum_{i} \alpha_{i}|i\rangle_{d i r}\left|\psi_{i}\right\rangle_{d a t}
\end{equation}
The reason why we introduce qRAM here is that, in this work, it plays an important role in the preparation and observation of the quantum state.

{\bf Preparation and observation of quantum state.}
The preparation and the observation of the quantum state are two inevitable problems. They are also important for the design of our quantum Fourier convolutional network, since our hybrid circuit needs to prepare and access the data structure in the quantum network. 
Thus, in this part, we also briefly introduce how to prepare the input quantum state on the basis of Theorem \ref{thm1}~\cite{moreno2020circuit}, and the pseudo-code about tomography observation in supplementary materials.
\newtheorem{thm}{\bf Theorem}[section]
\begin{thm}\label{thm1}
Let $M \in \mathbb{R}^{n^{\prime} \times n^{\prime}}$ be a matrix. Suppose $w$ is the number of nonzero entries, we can access a quantum data structure with size $O\left(w \log ^{2}\left(n^{\prime 2}\right)\right)$, which takes time $O\left(\log \left(n^{\prime 2}\right)\right)$ to store or update an entry. And quantum algorithms with access to this data structure can perform the following maps to precision $\epsilon^{-1}$ in time $O\left(p o l y \log \left(n^{\prime} / \epsilon\right)\right)$:
\begin{equation}
\begin{array}{c}
U_{\mathcal{M}}:|i\rangle|0\rangle \rightarrow \frac{1}{\left\|M_{i} .\right\|} \sum_{j} M_{i j}|i j\rangle \\
U_{\mathcal{N}}:|0\rangle|j\rangle \rightarrow \frac{1}{\|M\|_{F}} \sum_{i}\left\|M_{i}.\right\||i j\rangle
\end{array}
\end{equation}
where $\left\|M_{i \cdot} \right\|$ is the $l_{2}-$norm of the row i in $M$. This means that we can prepare an quantum approximation, i.e, $1 /\|v\|_{2} \sum_{i} v_{i}|i\rangle$, which is $\epsilon-$close to given classical vector in time $O\left(p o l y \log \left(n^{\prime} / \epsilon\right)\right)$.
\end{thm}

Based on the above theorem and qRAM, we can map the initial quantum state $|0\rangle^{\otimes N}$ to the quantum state with the amplitudes that we need quickly.

As for observation, suppose the input kernel size is $n \times n$,  we need to measure the transformed kernel firstly, and then observe the output state with $|x\rangle = \sum_{i \in[n^{2}]} x_{i}|i\rangle$, where $\|x\|=1$. In our hybrid circuit, we need to transform $|x\rangle$ to classical vectors for the subsequent calculations and optimization. Here we amend tomography algorithm in work \cite{kerenidis2019quantum}, since our result has its independent probability and parameters.

\section{Quantum Fourier Convolutional Network}

To achieve potential quantum speedups of the traditional CNNs, in this section, we propose a quantum Fourier convolutional network (QFCN). The framework of our QFCN is illustrated in Figure \ref{fig:qfcnn}. We replace the traditional convolutional layers with our Fourier convolutional layers in a quantum way, and utilize the algorithms introduced in section \ref{section:preliminary} for preparing and observing the quantum input and output. Below we introduce the quantum Fourier convolutional layer and the quantum back-propagation method for our QFCN model in detail.

\subsection{Quantum Fourier Convolutional Layer}
Assume at the layer $l$ of the CNN, the input is the feature maps (or image): $X^{\ell} \in \mathbb{R}^{H_{\ell} \times W_{\ell} \times C_{\ell}}$, where $H_{\ell}$, $W_{\ell}$, and $C_{\ell}$ are the height, width, and channel number, respectively.
Let $K^{\ell} \in \mathbb{R}^{h_{\ell} \times w_{\ell} \times C_{\ell} \times C_{\ell+1}}$ denote the convolutional kernels. 
Hence, at layer $\ell$, the convolution can be formulated as:
\begin{equation}
X_{i_{\ell+1}, j_{\ell+1}, c_{\ell+1}}^{\ell+1}=\sum_{i=0}^{h_{\ell}} \sum_{j=0}^{w_{\ell}} \sum_{c=0}^{C_{\ell}} X_{i+i_{\ell+1}, j+j_{\ell+1}, c}^{\ell} K_{i, j, c, c_{\ell+1}}^{\ell}  
\end{equation}
where the slice $\left(i_{\ell+1}, j_{\ell+1}, c_{\ell+1}\right)$ indicates one element of $X^{\ell+1}$.

\begin{figure}[t]
\begin{center}
{\includegraphics[width=1\linewidth,trim={0cm 3cm 2cm 0cm},clip]{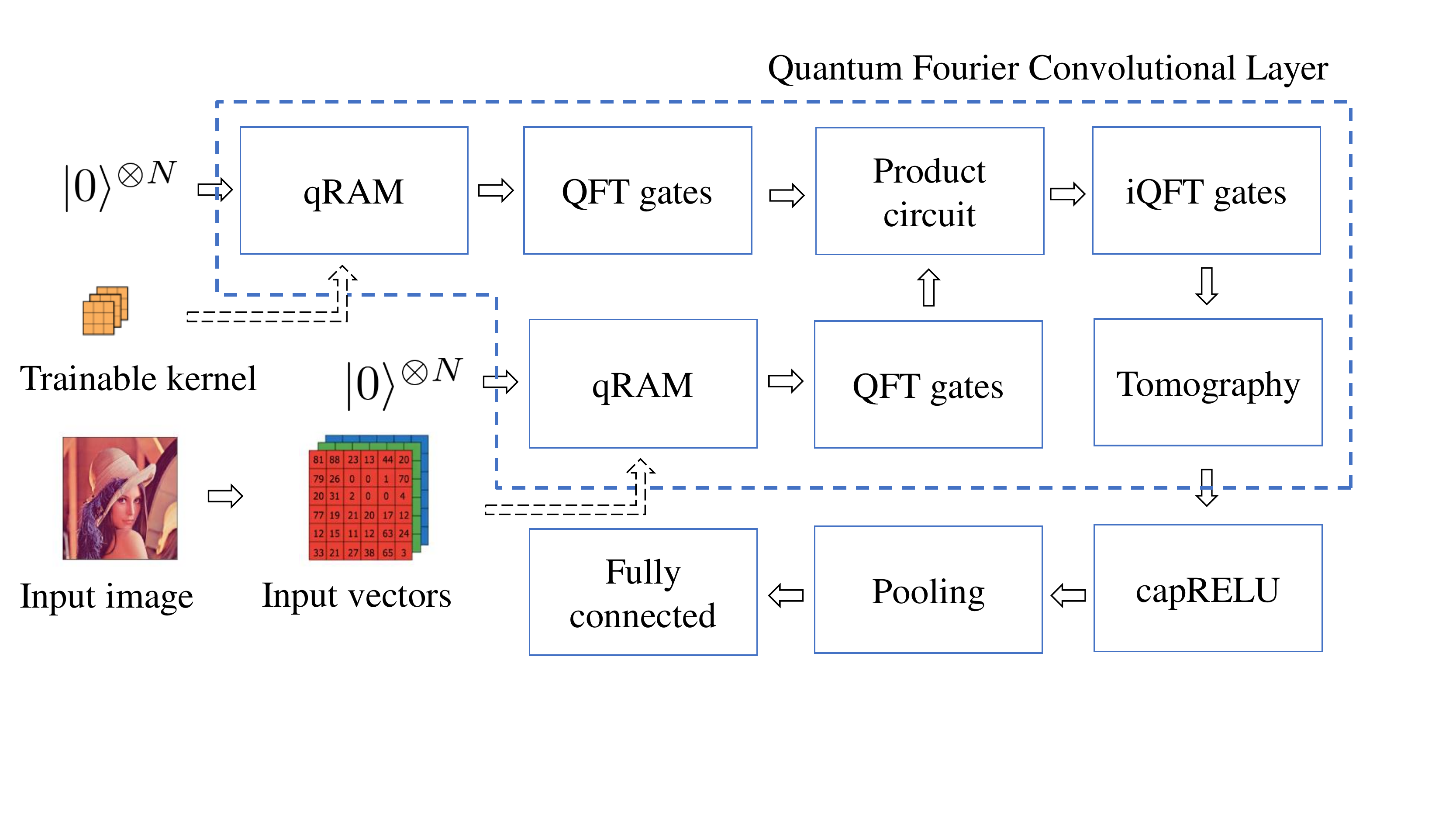}}
\end{center}
   \caption{Illustration of our QFCN with a quantum Fourier convolutional layer. 
   }
\label{fig:qfcnn}
\end{figure}


Based on the convolution theorem, the convolutional operation can be accomplished by transforming the kernel and data to Fourier basis, and then performing inverse transform based on their product.
Here to achieve quantum speedups for the classical convolutional layer, we need to replace its classical Fourier transform part with quantum Fourier transform and find the process $P$ to achieve product on combination of quantum state, i.e., $X^{\ell}*K^{\ell}=iQFT\left(P\left(QFT\left( K^{\ell} \right)\otimes QFT\left( X^{\ell} \right) \right) \right)$.
Unfortunately, there is no such a process $P$ that is able to complete the convolution (as suggested by \cite{lomont2003quantum}), as follows:


\begin{equation}
\sum_{t, j=0}^{N-1} \hat{f}(t) \hat{g}(j)|t\rangle|j\rangle \stackrel{P}{\longrightarrow} \lambda \sum_{k=0}^{N-1} \sum_{j=0}^{N-1} \hat{f}(j) \hat{g}(k-j)|k\rangle
\end{equation}
where $\lambda$ is a normalization factor. $\sum_{t, j=0}^{N-1} \hat{f}(t) \hat{g}(j)|t\rangle|j\rangle$ is the combination of the quantum-transformed kernel and quantum-transformed input. Previous works \cite{kerenidis2019quantum} 
attempt matrix product 
to circumvent this challenge. 
Here focusing on Fourier transform, we handle this issue and implement the quantum Fourier convolutional layer as follows:

\



\begin{equation}
X^{\ell+1}=iQFT\left(M^{QFT\left( K^{\ell} \right)} \left(QFT\left( X^{\ell} \right) \right) \right)
\end{equation}
where $QFT$ means quantum Fourier transform, and $iQFT$ means inverse quantum Fourier transform. $M\left( \cdot \right)$ performs the following mapping:
\begin{equation}
\sum_{t, j=0}^{N-1} \hat{f}(t) \hat{g}(j)|t\rangle|j\rangle \stackrel{M}{\longrightarrow} \sum_{t, j=0}^{N-1} \hat{f}(t) \hat{g}(t+j)|t\rangle|j\rangle
\end{equation} 

Inspired by \cite{ma2019windowed}, after applying inverse quantum Fourier transform on $\sum_{t, j=0}^{N-1} \hat{f}(t) \hat{g}(t+j)|t\rangle|j\rangle$, we get the convolution of $\sum_{k=0}^{N-1} f(k)|k\rangle$ and $\sum_{k=0}^{N-1} g(k) e^{-2 \pi i k j}|k\rangle$ in circuit of input. Since the measurement $j$ in circuit of kernel can be zero with probability $\sum_{k=0}^{N-1}|\hat{f}(k) \hat{g}(k)|^{2}$, we can 
get the convolution of $\sum_{k=0}^{N-1} f(k)|k\rangle$ and $\sum_{k=0}^{N-1} g(k)|k\rangle$ (quantum states of our kernels and input) when $j = 0$ is observed~\cite{ma2019windowed}.

Based on the above mechanism, we can construct the quantum Fourier convolutional layer as illustrated by the dashed box in Figure \ref{fig:qfcnn}.
Below we also present more details of the quantum Fourier convolutional layer, including revisiting the discrete and fast Fourier transform, and introducing the quantum Fourier transform (QFT).


\subsubsection{Discrete and Fast Fourier Transform}
\label{section:fft}
The Discrete Fourier Transform (DFT) is widely used in digital signal processing and image processing \cite{winograd1978computing}. It can map a vector $x=\left(x_{0}, x_{1}, \ldots, x_{N-1}\right)^{T}$ to a new vector $y=\left(y_{0}, y_{1}, \ldots, y_{N-1}\right)^{T}$, as follows:
\begin{equation}
y_{k}=\frac{1}{\sqrt{N}} \sum_{j=0}^{N-1} x_{j} e^{2 \pi i \frac{k j}{N}}
\label{formula1}
\end{equation}

The Fast Fourier transform (FFT) \cite{nussbaumer1981fast}  
is for fast computation of the DFT,
which makes use of the symmetries in the DFT and can reduce the computational cost from $\mathcal{O}\left[N^{2}\right]$ to $\mathcal{O}[N \log N]$. 

\subsubsection{Quantum Fourier Transform}
The quantum Fourier transform (QFT) has the same input and output as the DFT, but QFT is for the quantum state and can achieve exponential speedup. It accepts a quantum state $|\psi\rangle=\sum_{j=0}^{N-1} x_{j}|j\rangle$ as input, and performs the following mapping on its quantum circuit:
\begin{equation}
\sum_{j=0}^{N-1} x_{j}|j\rangle \stackrel{QFT}{\longrightarrow} \sum_{k=0}^{N-1} y_{k}|k\rangle
\label{formula2}
\end{equation}  
where the amplitude $y_{k}$ is the same as Eq. \eqref{formula1}. By substituting Eq. \eqref{formula1} into Eq. \eqref{formula2} and offsetting the amplitudes $x_{j}$, it can be written as:
\begin{equation}
|j\rangle \rightarrow \frac{1}{\sqrt{N}} \sum_{k=0}^{N-1} e^{2 \pi i \frac{k j}{N}}|k\rangle
\label{formula3}
\end{equation}

\begin{figure}[t]
\begin{center}
{\includegraphics[width=\linewidth,trim={0cm 4.6cm 0cm 1.2cm},clip]{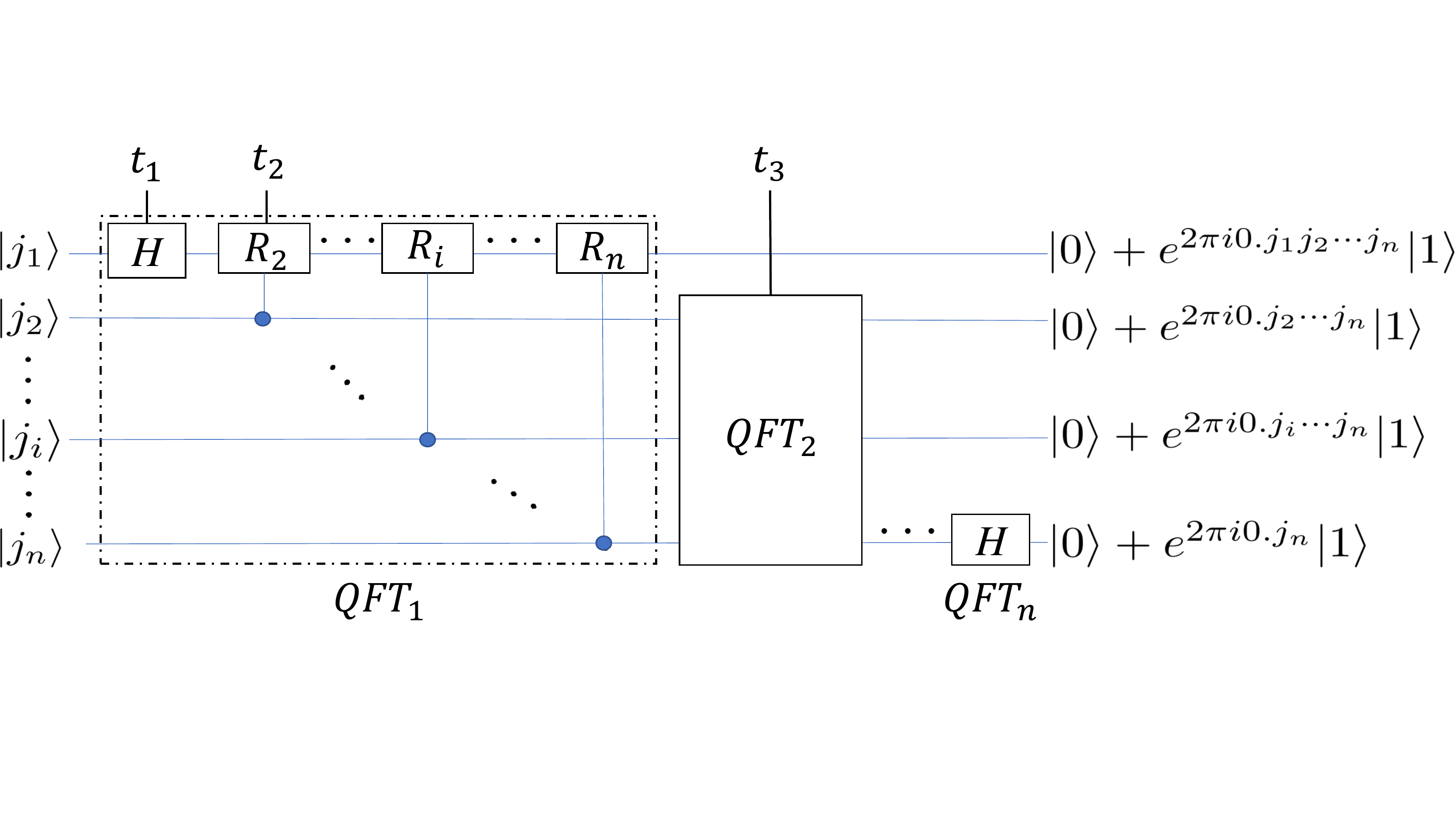}}
\end{center}
   \caption{Circuit for QFT gates without swapping gate}
\label{fig:qft}
\label{fig:onecol}
\end{figure}

It is obvious that the data we need is on the amplitudes of the quantum state. Then we describe what happens on the input $|\psi\rangle$, and analyse why QFT can be faster than FFT by the circuit for the QFT illustrated in Figure \ref{fig:qft}.
The quantum controlled gate $R_{k}$ denotes the unitary operator:

\begin{equation}
R_{k} =\left[\begin{array}{cc}1 & 0 \\ 0 & e^{2 \pi i / 2^{k}}\end{array}\right]
\end{equation}

After time $t_{1}$, the Hadamard gate is applied on the first qubit of the state, which produces:

\begin{equation}
\dfrac{1}{2^{1 / 2}}\left(|0\rangle+e^{2 \pi i 0 . j_{1}}|1\rangle\right)\left|j_{2} \ldots j_{n}\right\rangle
\end{equation}

According to the Euler formula $e^{ \pi i }=-1$, we get:

\begin{equation}
%
e^{2 \pi i 0 . j_{1}} = 
\left\{
             \begin{array}{lr}
             -1, ~~ j_{1}=1  \\
             
             ~~~1, ~~ otherwise
             \end{array}
\right.
\end{equation}

Applying the $controlled-R_{2}$ gate on the first qubit, we have the state in time $t_{2}$:

\begin{equation}
\frac{1}{2^{1 / 2}}\left(|0\rangle+e^{2 \pi i 0 . j_{1} j_{2}}|1\rangle\right)\left|j_{2} \ldots j_{n}\right\rangle
\end{equation}

We repeatedly apply $controlled-R_{i}$ gate on the first qubit. After $controlled-R_{n}$ gate, we have the state:

\begin{equation}
\frac{1}{2^{1 / 2}}\left(|0\rangle+e^{2 \pi i 0 . j_{1} j_{2} \ldots j_{n}}|1\rangle\right)\left|j_{2} \ldots j_{n}\right\rangle
\end{equation}

The second qubit is the same as the first one, and we get the state after $t_{3}$:
\begin{equation}
\begin{array}{l}
\frac{1}{2^{2 / 2}}\left(|0\rangle+e^{2 \pi i 0 . j_{1} j_{2} \ldots j_{n}}|1\rangle\right)\left(|0\rangle+e^{2 \pi i 0 . j_{2} \ldots j_{n}}|1\rangle\right) \vspace{1ex}\\
\left|j_{3} \ldots j_{n}\right\rangle
\end{array}
\end{equation}

Here we can also see the robustness of the QFT circuit. We do not need to prepare a new circuit to adapt the length of the quantum state. The whole circuit can be regarded as the assemble of several separate QFT gates, while the single QFT gate is made by several controlled$-R_{k}$ gates and a Hadamard gate to ensure that the QFT gate is unitary.

After applying $n$ QFT gates on $n$ qubits, we swap the result and get the state:

\begin{equation}\begin{array}{l}
\dfrac{1}{2^{n / 2}}\left(|0\rangle+e^{2 \pi i 0 . j_{n}}|1\rangle\right)\vspace{1ex}\\\left(|0\rangle+e^{2 \pi i 0 . j_{n-1} j_{n}}|1\rangle\right)\ldots\left(|0\rangle+e^{2 \pi i 0 . j_{1} j_{2} \cdots j_{n}}|1\rangle\right)
\end{array}\end{equation}
 This quantum state is equivalent to the quantum state in Eq. \eqref{formula3}, after being simplified by Euler formula.

As for the inverse QFT (iQFT) circuit, it can be achieved just like what it spells: reversing the circuit for the QFT~\cite{nielsen2002quantum}.

One of the motivations we exploit QFT in this paper is that the QFT algorithm can achieve exponential speed-up. As we can see, a total of $n+(n-1)+\cdots+1=n(n+1) / 2$ gates are required to perform the QFT on $n$ qubits. These $n$ qubits need $2^{n}$ bits to be encoded in classical computer and we need $n 2^{n}$ gates to perform FFT on these bits. Based on the number of gates, the QFT algorithm reduces time complexity from $\mathcal{O}\left(n 2^{n}\right)$ to $\mathcal{O}\left( n^{2}\right)$. The best approximation is able to reduce this complexity to $\mathcal{O}(n \log n)$~\cite{hales2000improved}.  

\subsection{Quantum Back-Propagation}
The optimization of the model has always been an essential part of quantum machine learning \cite{farhi2014quantum,hadfield2019quantum}. Previous works about quantum convolution neural network utilize quantum matrix product process and QEC (a mechanism to detect and correct local quantum errors) respectively to achieve optimization of their models \cite{cong2019quantum,kerenidis2019quantum}. Differently, in this work, we propose our optimization method using the advantages of the hybrid quantum-classical algorithm mentioned before.

Specifically, this 
QFCN model can still be trained with back-propagation like the classical CNNs. The quantum Fourier convolutional layer in the hybrid quantum-classical circuit can be seen as a quantum circuit with parametric gates (PQC). This means the gradients of our quantum circuit can be estimated by shifting the parameters and running the same QFT circuit~\cite{zhao2019qdnn}. Once we get the gradients, they can be incorporated into the classical back-propagation process using gradient-based algorithms \cite{rumelhart1986learning}. Denote $U(\vec{\theta})=U_{n}\left(\theta_{n}\right) \ldots U_{1}\left(\theta_{1}\right)$ as our PQC mathematical expression, and $H$ as the Hermitian operator representing our goal. Then the expectation quantum value should be $L(\vec{\theta})=\left\langle\psi_{0}\left|U^{\dagger}(\vec{\theta}) H U(\vec{\theta})\right| \psi_{0}\right\rangle$. 

Through a series of calculation and offsetting \cite{zhao2019qdnn}, the gradient of $j^{th}$ parameter is below:
\begin{equation}
\frac{\partial L}{\partial \theta_{j}}=\frac{1}{2}\left[L_{j,+}(\vec{\theta})-L_{j,-}(\vec{\theta})\right]
\end{equation}
where
\begin{equation}
\begin{array}{l}
L_{j,+}(\vec{\theta})=L\left(\theta_{1}, \ldots, \theta_{j-1}, \theta_{j}+\frac{\pi}{2}, \theta_{j+1}, \ldots, \theta_{n}\right) \\
L_{j,-}(\vec{\theta})=L\left(\theta_{1}, \ldots, \theta_{j-1}, \theta_{j}-\frac{\pi}{2}, \theta_{j+1}, \ldots, \theta_{n}\right)
\end{array}
\end{equation}
which demonstrates that the gradient can be accessed through shifting parameter in qRAM and run the same circuit.

Note that in some scenarios, we may need to stack multiply layers in the model, which may lead gradient to Barren plateaus~\cite{mcclean2018barren} and vanish, while the robustness of the QFT circuit that we mentioned above can help to achieve the layer-wise algorithm~\cite{skolik2020layerwise} and mitigate Barren plateaus. 

\section{Experiments}
In our experiments, we evaluate our proposed QFCN on two tasks.
First, we apply QFCN on image classification by following QCNN \cite{kerenidis2019quantum}. 
Then, we test QFCN on the traffic prediction task by inserting the quantum Fourier convolutional layer into the the spatio-temporal graph convolutional network. 

\subsection{Image Classification}

In this experiment, our QFCN is designed to accept the 2D image data structure. Specifically, we construct our QFCN using the same baseline CNN as \cite{kerenidis2019quantum}.


We apply QFCN on the MNIST dataset including 60,000 training images and 10,000 testing images. Each image presents a hand-written number consisting of 28$\times$28 grayscale pixels. Following the work of QCNN \cite{kerenidis2019quantum}, we use the PyTorch library to simulate the quantum environment by several parameters of noises. 
Similar to \cite{kerenidis2019quantum}, the cap of the non-linearity function $C$ is set to 10, 
and the quantum noise $\epsilon$ during preparation and observation is set to 0.01. Readers are referred to \cite{kerenidis2019quantum} for more details about the quantum simulation process. 
In this work, we use the same image samples as \cite{kerenidis2019quantum} to plot the curves.
As the input and output values of QFT and FFT will be the same, 
we perform simulation experiments by applying the aforementioned noise parameters to the process of imitating QFT with FFT,  
since such simulation settings are also adopted by previous works \cite{kerenidis2019quantum} owing to the inaccessibility of the real quantum computers.
Considering the GPU speedups of CNN, we do not compare the computation time of QFCN with the traditional CNN. But we have also conducted a toy experiment to directly compare the FFT with the classical convolution. In our simple test using an i5-6300HQ CPU with 2.30GHz, where the kernel size is 2$\times$3$\times$5$\times$5 and the signal size is 3$\times$1024$\times$1024, the FFT-based convolution used $744ms$ for calculation, which is faster than the classical convolution using $1.05s$. This situation is not really uncommon in some recent works \cite{chitsaz2020acceleration,li2020falcon} because of their large kernel size, and we should also note that theoretically, the QFT algorithm can be exponentially faster than the FFT algorithm.

\begin{figure}[t]
\begin{center}
{\includegraphics[width=1\linewidth]{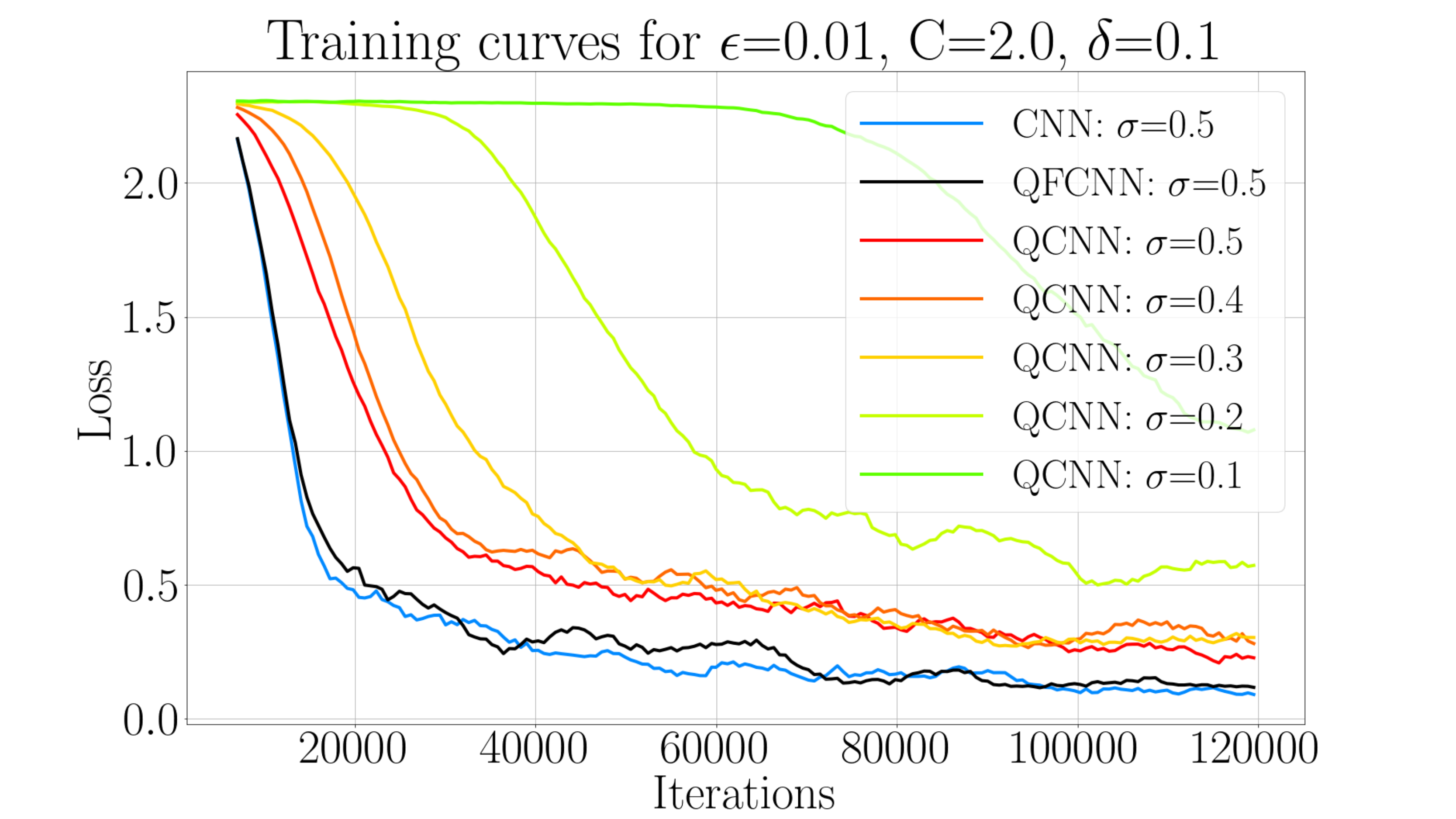}}
\end{center}
   \caption{Numerical simulations of the training of the QCNN \cite{kerenidis2019quantum} and our QFCN. }


\label{fig:curve}
\label{fig:onecol}
\end{figure}

\textbf{Result.}
We plot the convergence curves of the classical CNN, QCNN \cite{kerenidis2019quantum}, and QFCN in Figure \ref{fig:curve}. We do not show many more training curves for QCNN \cite{kerenidis2019quantum}, because QCNN \cite{kerenidis2019quantum} does not perform well under other parameters, as shown by the paper of \cite{kerenidis2019quantum}. We can also observe that our QFCN achieves a faster and easier convergence on the MINST dataset, compared to the QCNN. 

In Table \ref{tab:results}, we show the performance of our model, and compare it with the QCNN model. We can observe that our model achieves higher accuracy than QCNN. The performance disparity can be partially due to that our hybrid quantum-classical circuit avoids quantum sampling and quantum pooling, which are simulated by adding quantum noises in QCNN.
\begin{table}
\caption{Loss and accuracy of QFCN and QCNN. 
}
\begin{center}
	\scalebox{0.8}{
        \begin{tabular}{cc|cc}
        \hline
          \multicolumn{2}{c}{MNIST} & 
         Loss & Accuracy \\ \hline
        \multicolumn{2}{c}{QCNN ($\sigma=0.1$)} & 1.07 & 86.1  \\
        \multicolumn{2}{c}{QCNN ($\sigma=0.2$)} & 0.552 & 92.8  \\
		\multicolumn{2}{c}{QCNN ($\sigma=0.3$)} & 0.391 & 94.3  \\
		\multicolumn{2}{c}{QCNN ($\sigma=0.4$)} & 0.327 & 94.4  \\
        \multicolumn{2}{c}{QCNN ($\sigma=0.5$)} & 0.163 & 95.9 \\ \hline 
        \multicolumn{2}{c}{QFCN}& \textbf{0.113} & \textbf{96.3\%}  \\ \hline
        \end{tabular}
        }
        
\end{center}
\label{tab:results}
\end{table}

\subsection{Traffic Prediction}
Many existing works \cite{zhao2020go,ghosh2020stacked} use the models constructed with both temporal convolution and spatial graph convolution. This kind of networks is generally called Spatio-Temporal Graph Convolutional Network (STGCN) \cite{wu2020comprehensive,li2017diffusion,guo2019attention}. 
We apply our quantum Fourier convolution for the STGCN \cite{yu2017spatio} and use the designed quantum Fourier convolutional layer to handle the convolution for the time series information. In this experiment, we mainly aim to demonstrate the ability of our quantum Fourier convolutional layer for replacing the temporal convolution over large-scale series input in the baseline STGCN \cite{yu2017spatio}. We name this model as quantum Fourier Spatio-Temporal Graph Convolutional Network (qf-STGCN).

\begin{figure}[t]
\begin{center}
{\includegraphics[width=1\linewidth]{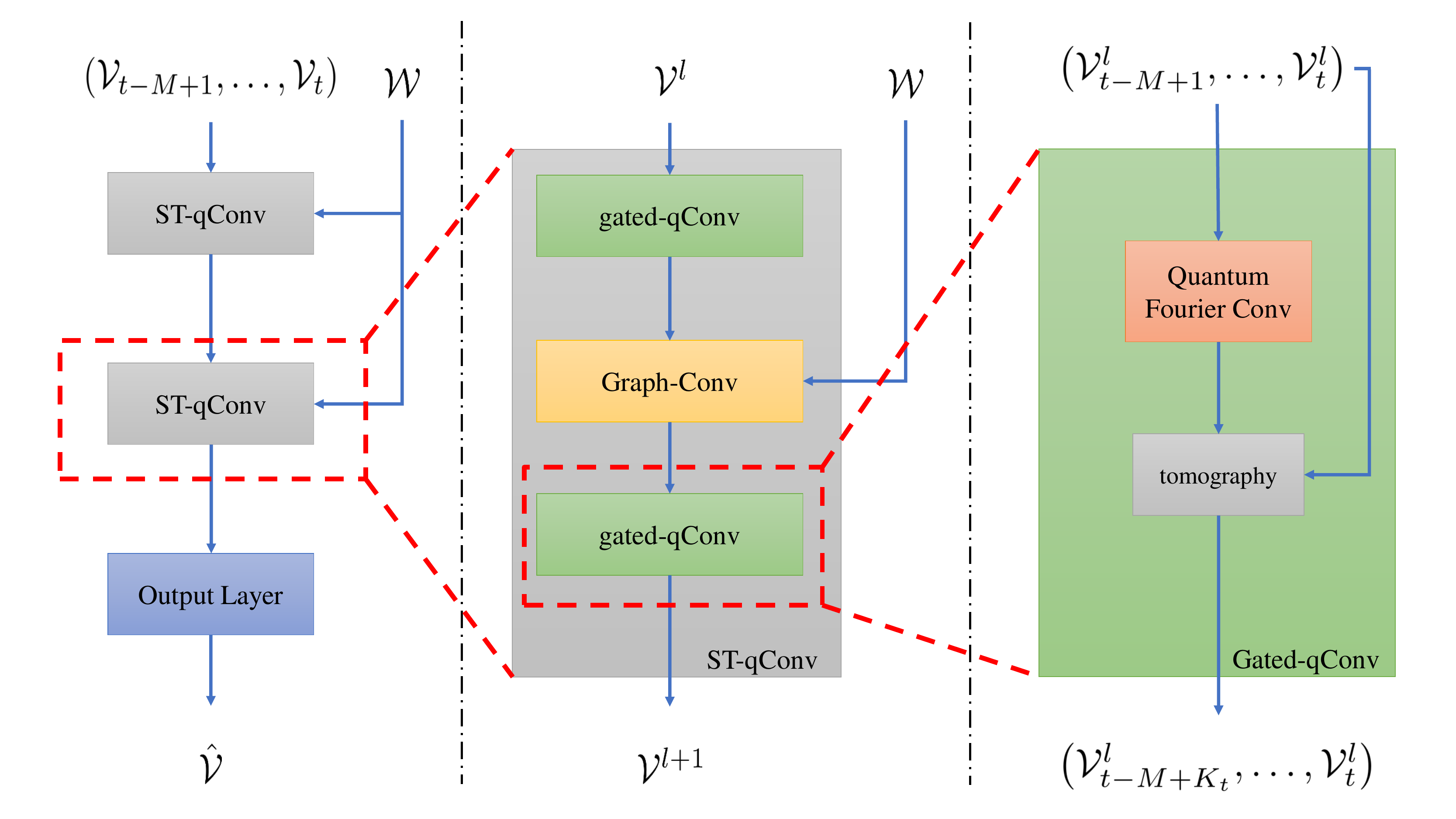}}
\end{center}
   \caption{The architecture of the qf-STGCN. The input $\mathcal{W}$ denotes the weighted adjacency matrix of the graph. One single $\mathcal{V}_{t}$ represents the data captured at time $t$. $\hat{\mathcal{V}}$ denotes the final prediction, and $\mathcal{V}^{l+1}$ denotes the input data of layer $l+1$. $K_{t}$ is the width of the kernel. $M$ is the length of the input series.
   The qf-STGCN is adapted from the classical STGCN in \cite{yu2017spatio}. Refer to \cite{yu2017spatio} for more details of the original STGCN.
   }
\label{fig:qstgcn}
\end{figure}

The qf-STGCN uses the quantum mechanism to predict time series on the basis of previous time series and graph-structured data. We adopt most of the structure parts of the previous STGCN \cite{yu2017spatio}. Specifically, the spatio-temporal convolutional blocks in STGCN contain two gated-convolutional blocks for handling temporal properties, and one graph convolutional block to handle spatial properties of the graph. We utilize our quantum Fourier convolutional layer to replace the gated convolutional layer, and add simulation of tomography on the quantum machine. The framework of qf-STGCN is illustrated in Figure \ref{fig:qstgcn}. 
 
We assess the performance of the model with two metrics, namely, Mean Absolute Error (MAE) and Mean Squared Error (MSE), following the previous works on traffic prediction  \cite{li2017diffusion,seo2018structured}, where $\textbf{MAE} ={\frac{1}{n}}{\sum_{t=1}^{n}|{{\mathcal{V}_{t}}}-\hat{\mathcal{V}_{t}}|}$, 
and
$\textbf{MSE} =  {\frac{1}{n}}{\sum_{t=1}^{n}}{({{\mathcal{V}_{t}}}-\hat{\mathcal{V}_{t}})}^2$. $\mathcal{V}_{t}$ and $\hat{\mathcal{V}_{t}}$ are the ground truth and prediction of traffic condition in time $t$ respectively.

We adopt the METR-LA dataset \cite{li2017diffusion} that is widely used in traffic prediction works. The METR-LA is a dataset about information collected by loop sensors in the highway of Los Angeles, which contains 207 nodes, 1515 edges and 34272 time steps. The distribution of the sensors can be found in Li \cite{li2017diffusion}. 

In this experiment, we add a baseline which puts our qf-STGCN on the pure quantum circuit with numerical simulations of quantum sampling, quantum pooling, and quantum back-propagation, as done by \cite{kerenidis2019quantum}.

\begin{figure}[t]
\begin{center}
{\includegraphics[width=1\linewidth]{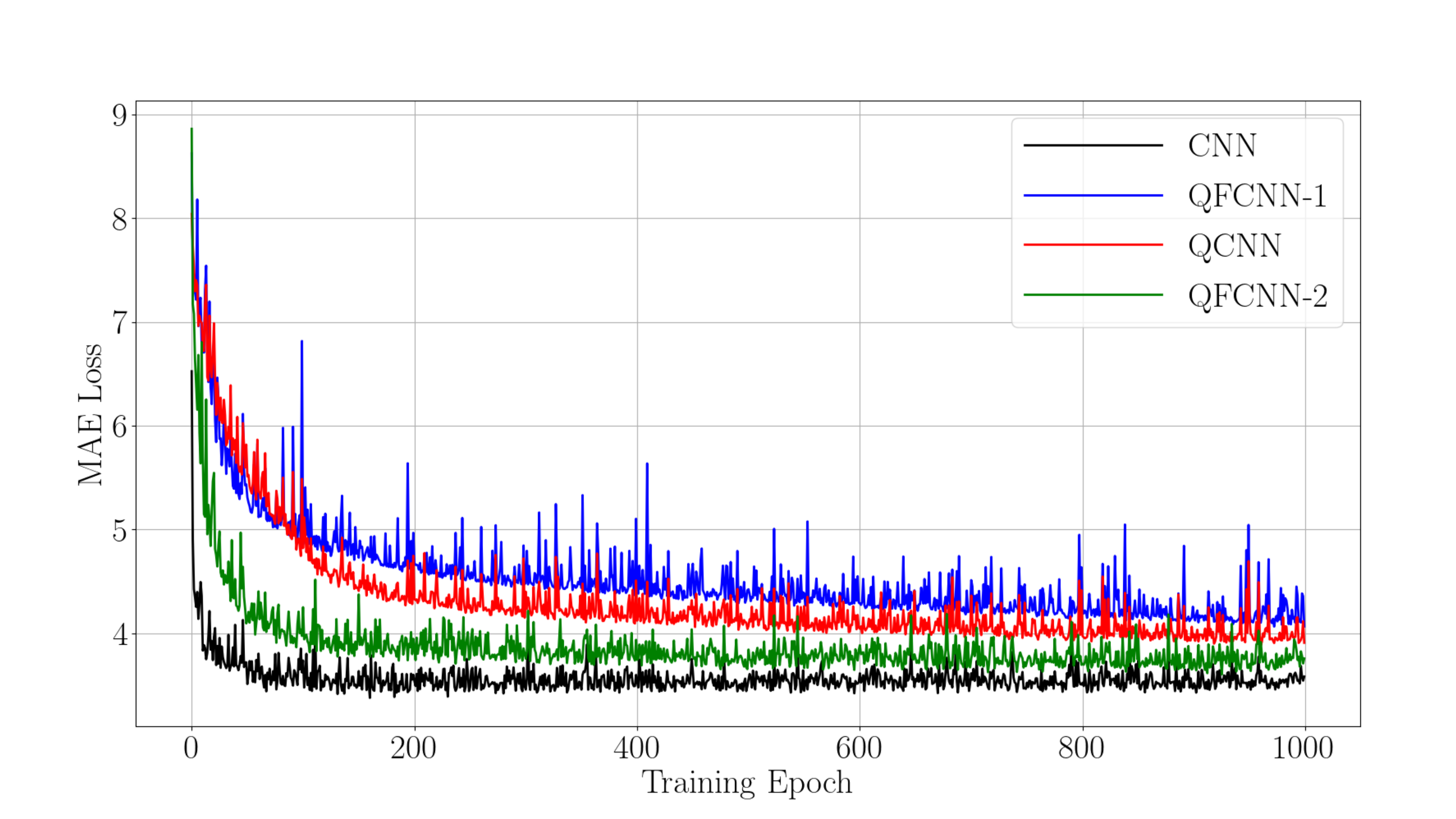}}
{\includegraphics[width=1\linewidth]{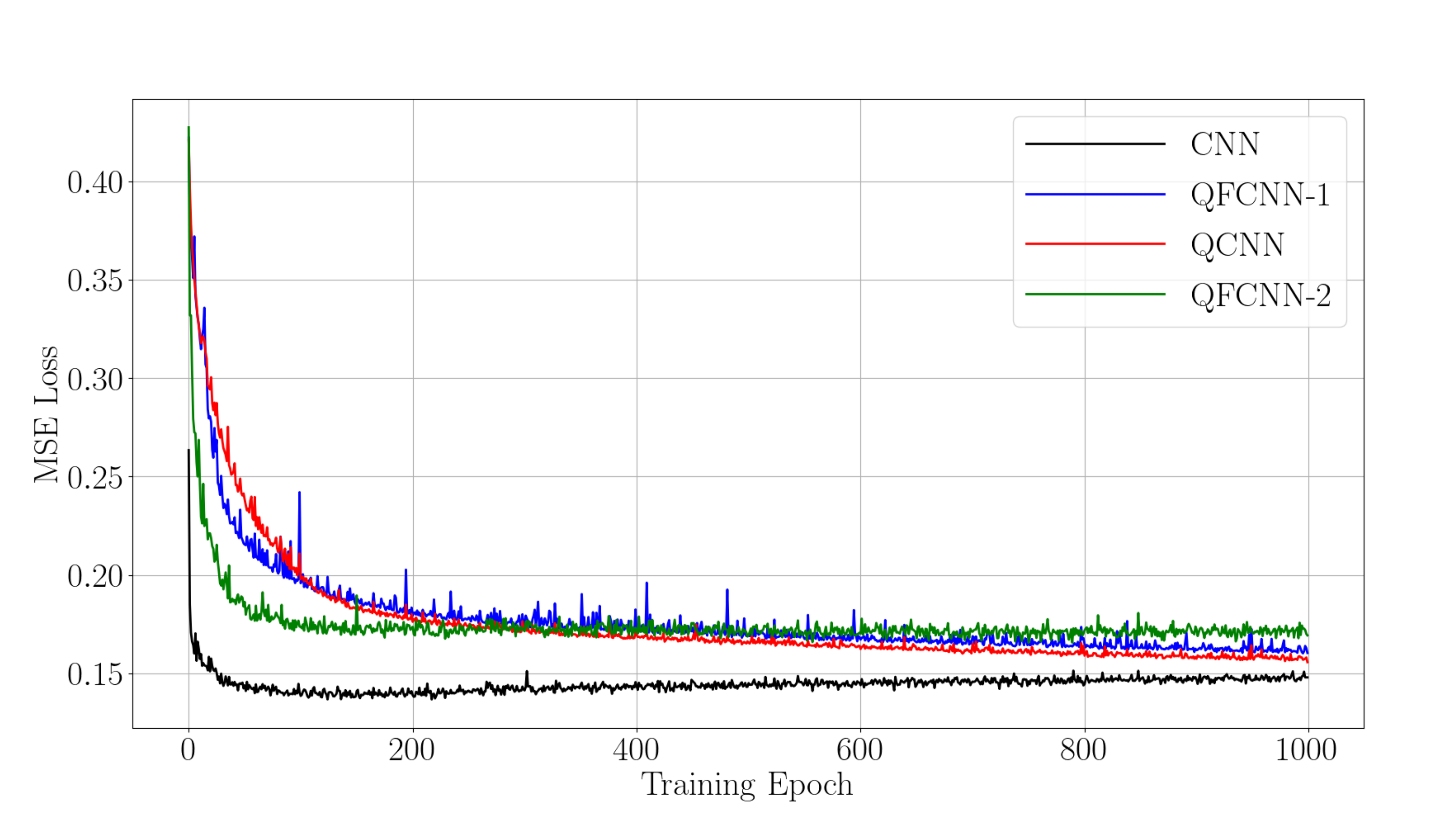}}   
\end{center}
   \caption{The MAE and MSE curves of STGCN under different settings. The blue curve represents the qf-STGCN on the hybrid circuit, named q-STGCN-1. The red curve indicates the qf-STGCN in the pure quantum environment, named q-STGCN-2. The black curve represents the performance of the origin STGCN on the same dataset.}
\label{fig:error}
\label{fig:onecol}
\end{figure}

\textbf{Result.}
We can observe that in Figure \ref{fig:error}, both of our qf-STGCN models can still converge fast and correctly. However, the data structure of the spatial-temporal graph in this task does not have many parts of redundant information like images. Besides, some previous works have also found that GCN is vulnerable to attacks and noise \cite{dai2018adversarial,zugner2018adversarial}, and then when we change the attribution on the node, the performance of the following graph convolution will also be influenced. This implies that there could be a need of more robust quantum graph convolution~\cite{verdon2019quantum}. Nevertheless, despite of the influences of quantum noise on the graph data, our model can still be used for quantum speedups. This experiment also shows the ability of our model for handling the large-scale time series dataset. 

\section{Conclusion}
Quantum computation and CNN have become two popular research fields recently. In this work, we have presented a new quantum Fourier convolutional network. Our QFCN could be able to run on both hybrid quantum-classical circuits and pure quantum computers. Our experiments show our model's ability of embedding within the classical CNNs and handling large datasets. This means our model can also be inserted into many other neural networks with convolutional layers to achieve quantum speedups. There are also some shortcomings: like many other quantum machine learning work, we still lack of a real programmable quantum computer to prove our research and we believe that day is coming soon. 
{\small
\bibliographystyle{unsrt}  
\bibliography{references}
}

\end{document}